# Inter-residue, inter-protein and inter-family coevolution: bridging the scales

Hendrik Szurmant[1+], Martin Weigt[2+]

**Affiliations**:

1. Department of Basic Medical Sciences, College of Osteopathic Medicine of the Pacific, Western University of Health Sciences, Pomona CA 91766, USA
2. Sorbonne Universités, UPMC Université Paris 06, CNRS, Biologie Computationnelle et Quantitative - Institut de Biologie Paris Seine, 75005 Paris, France

+ to whom correspondence might be addressed

**Corresponding authors**:

Hendrik Szurmant: 309 E. 2nd Street, Pomona, CA 91766-1854, USA, tel. +1-909-706-3938, email: hszurmant@westernu.edu

Martin Weigt: LCQB UMR7238, 4 place Jussieu 75005 Paris, France, tel. +33 1 44 27 73 68, email: martin.weigt@upmc.fr

**Short title:** Coevolutionary modeling of interacting proteins

## Abstract

Interacting proteins coevolve at multiple but interconnected scales, from the residue-residue over the protein-protein up to the family-family level. The recent accumulation of enormous amounts of sequence data allows for the development of novel, data-driven computational approaches. Notably, these approaches can bridge scales within a single statistical framework. While being currently applied mostly to isolated problems on single scales, their immense potential for an evolutionary informed, structural systems biology is steadily emerging.

## Highlights

- Coevolutionary modeling has an immense potential in structural systems biology.
- Interaction between two protein families can be detected.
- Specific interaction partners inside families can be deduced.
- Contact residues between interacting proteins can be predicted.
- Predicted inter-protein contacts can guide in silico protein-complex assembly.

**Introduction**

The importance of protein-protein interactions (PPI) in living systems is unquestioned. However, concerning PPI, different research communities are interested in differing aspects. Systems biologists are primarily interested in the complex networks formed by PPI, with proteins being the nodes, and links indicating interactions between two nodes [1]. Molecular or structural biologists will request more detailed information about the links in such networks [2] to gain insight into the interaction mechanism: how do these proteins interact, what are the interfaces, what does the protein complex look like? Evolutionary biologists may ask if a PPI present in one species exists in other species, to characterize the level of evolutionary conservation and innovation in PPI.

These scales play together in coevolutionary studies of interacting proteins. The evolutionary conservation of PPI across many different proteins of the same family is an essential ingredient: Coevolutionary PPI studies require large multiple-sequence alignments (MSA) of two homologous protein families, each one typically containing thousands of proteins from hundreds of diverged genomes. Employing these two MSA, one can ask three fundamental questions concerning different scales of PPI:
1. *Do the two protein families interact?* More precisely, do the two MSA contain a substantial number of protein pairs in different species, which interact [3]?
2. *Which protein pairs from the two MSA interact, specifically?* Even if we know that two families interact, we do not yet know, which specific protein pairs interact. An illustrative example is the two-component signal transduction systems (TCS) in the bacteria: bacterial genomes typically code for around 20 different TCS that connect differing signals with differing response. The vast majority of TCS employ the same two protein families – histidine sensor kinases (HK) and response regulators (RR). Interaction specificity and avoided cross talk are essential to obtain faithful responses to external signals [4,5].
3. *How do the proteins interact?* Can we identify the interaction interface? Can we predict residues being in contact across this interface? Can we assemble the complex from the individual protein structures?

In the last few years, significant progress in coevolutionary modelling [6,7] and the rapid growth of protein sequence databases [8,9] have allowed application to all three above areas of interest. It is now clear that sequences alone can be used to gain at least partial answers to all three questions, cf. Fig. 1 for a schematic overview.

These questions have been addressed in opposite order in the recent coevolution literature, from the finer residue scale over the protein scale up to the family scale. This opposite direction will be followed in this manuscript. To begin, we will review shortly the main arguments for the recently developed coevolutionary methods in the case of single proteins, more specifically in the prediction of native contacts between residue pairs using sequence information alone.

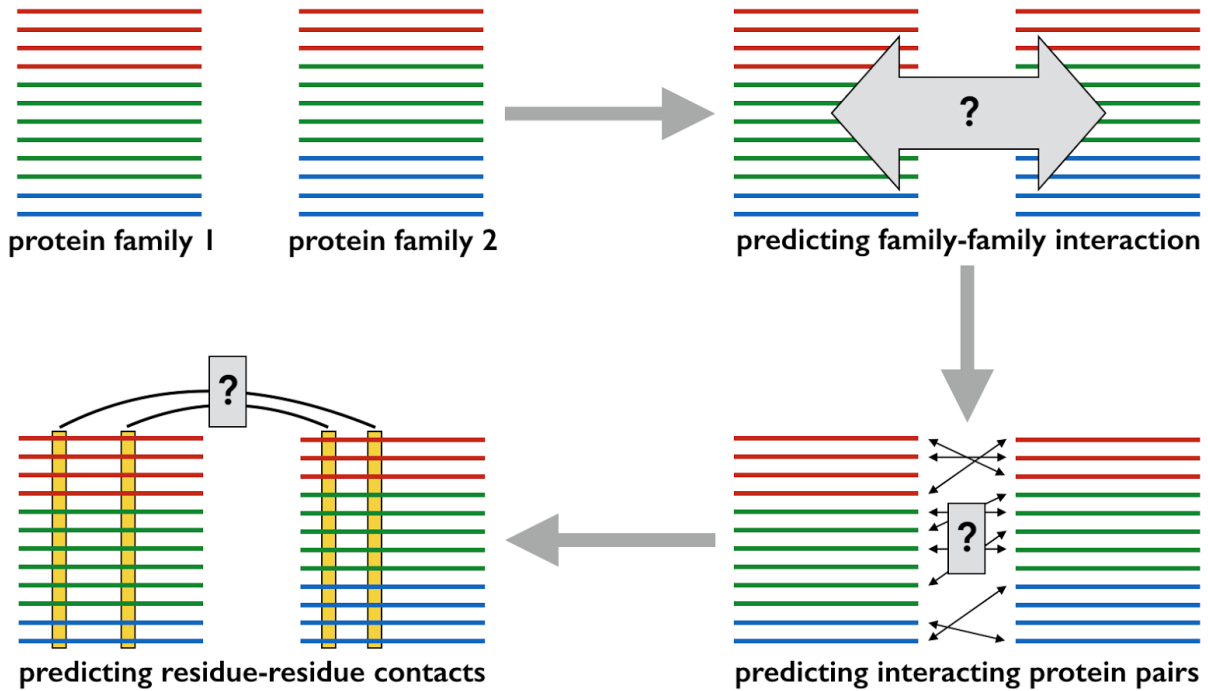

**Figure 1. Schematic overview over the multiple scales of PPI, which can be addressed by coevolutionary analysis** – Starting from two MSA of homologous protein families (upper left: paralogs inside the same species are symbolized by identical colors, different species by distinct colors), coevolutionary analysis addresses the following questions: (i) Do the two families interact (upper right)? (ii) If yes, which specific protein pairs do interact (lower right)? (iii) How do these proteins interact; which residues are in contact across the PPI interface (lower left)? Even if each question requires the previous one to be answered first, recent coevolutionary studies have approached them mostly in the opposite order, by using input derived from other approaches and not based on inter-protein coevolution.

The main idea underlying coevolutionary analysis is that most amino-acid substitutions in a protein perturb the physico-chemical coherence between the substituted residue and its neighbors in the three-dimensional (3D) protein structure. Coherence may be reestablished by compensatory substitutions in the contacting residues. This simple argument implies that residues in contact show correlations in their amino-acid occurrences, even if well separated along the primary sequence [10,11]. However, correlations remain of limited use for detecting residue-residue contacts: when residue $i$ coevolves with residue $j$ and residue $j$ with residue $k$, also $i$ and $k$ are correlated. Recent global coevolutionary methods, like DCA (direct coupling analysis) [12,13], PsiCov [14] or Gremlin [15,16], are able to disentangle such indirect correlations, and extract direct coevolutionary couplings. These have been found to accurately predict residue-residue contacts – provided a sufficiently large MSA. Predicted contacts have subsequently guided the prediction of hundreds of unknown protein structures [17-21]. The different mentioned methods show very similar performances; they are based on different approximations (mean field, Gaussian, pseudo-likelihood maximization) to achieve the same computational hard inference goal. For simplicity, we will mainly refer to all these methods as being of "DCA-type", and specify the specific method if needed.

**Coevolutionary analysis detects inter-protein residue-residue contacts**

Currently, the major application of DCA-type methods consists in tertiary protein structure prediction. Yet DCA has initially been proposed in the context of PPI [13]. The aim was to use the amino-acid covariation between the aforementioned HK and RR in bacterial TCS, to predict inter-protein residue-residue contacts and to assemble the – at that time – unknown protein complex (only the structurally similar but not homologous Spo0B/Spo0F complex was solved [22]). Limited power of mutual information (correlation) as contact predictor [23] has inspired the use of a global statistical model. This, via so-called direct couplings, collectively reproduces the amino-acid conservation and covariation statistics. Direct coevolutionary couplings, as derived by DCA, largely outperform mutual information or correlations in contact prediction, cf. Figs. 2A and 2B. Subsequent molecular simulations were guided by the predicted contacts, to assemble monomeric HK and RR into a quaternary structure model [24]. Compared to a crystal structure published in parallel [25], the model had a root-mean square deviation of only 3.6Å.

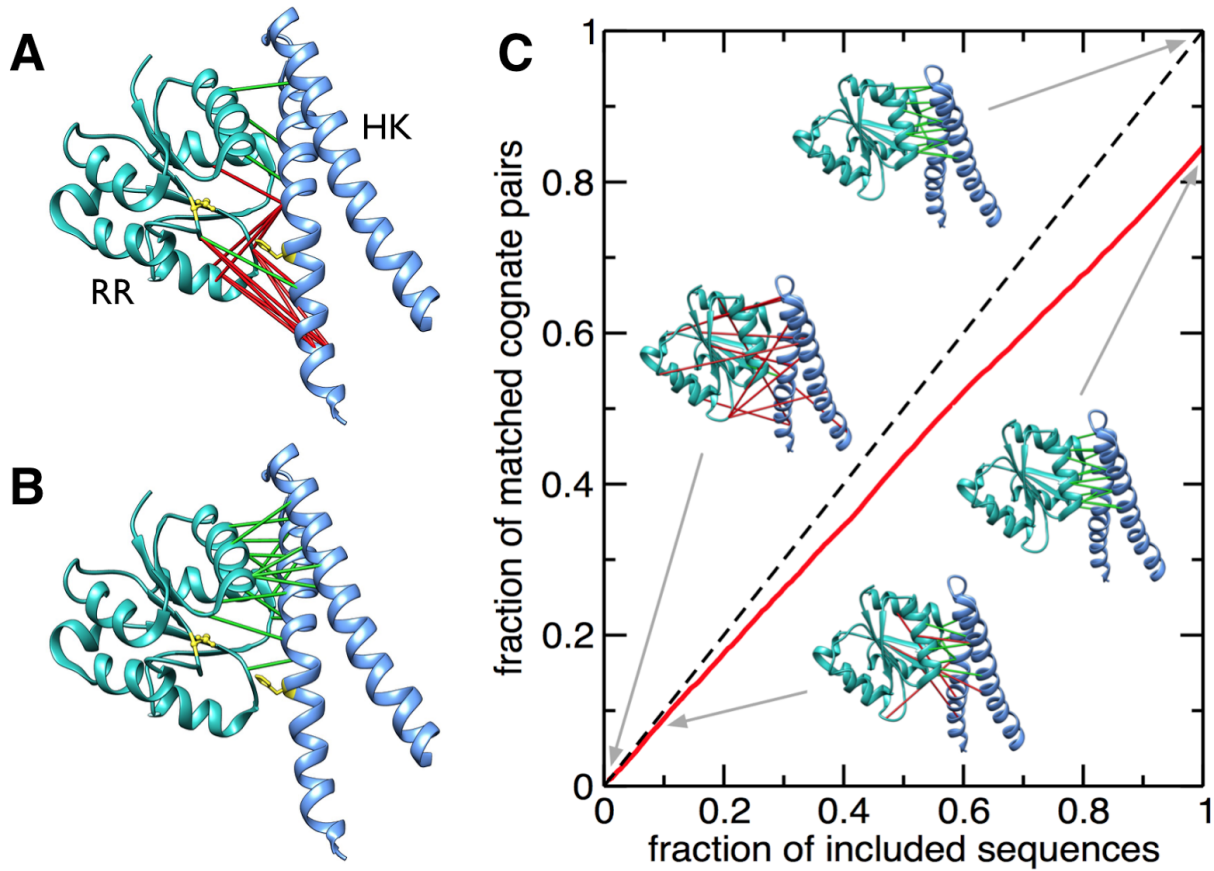

**Figure 2. Coevolutionary prediction of inter-protein residue-residue contacts and specific protein-protein interaction partners** – The left panels show the first 15 inter-protein residue-residue correlations (A) and direct couplings (B) for the HK/RR interaction (green – true positive, i.e. in contact across the interface, red – false positive, i.e. distant across the interface). While it is impossible to bring all predicted pairs in Panel A into simultaneous contact, Panel B suggests a quaternary structure bringing also the active sites (yellow) into spatial vicinity, as needed for phosphotransfer. Panel C (from [26]) shows the result of the progressive matching procedure between paralogs applied to 8,998 TCS from 712 species: the red line follows, from the lower left to the upper right corner, the progressive inclusion of more and more species into the paralog matching. It shows the fraction of correctly matched pairs (the so-called cognates) as a function of the currently matched sequences. While a perfect algorithm would follow the dashed diagonal (all included sequences are correctly matched), the algorithm of [26] finally matches 86% of all 8,998 pairs correctly. A random algorithm would correctly match, on average, one protein pair

per species, corresponding to only about 8% of all pairs. Contact predictions at different stages of the matching procedure are shown (green – true positive, red – false positive), for 59 sequences (seed MSA), for about 1000 sequences, for the fully matched MSA, and for the correct MSA of cognate pairs (cf. grey arrows). The quality of the contact prediction for the full matching is very close to the one used the correct cognate pairing.

Using the same ideas but more efficient algorithmic approaches, on the order of 100 PPI were analyzed and, lacking experimental structures assembled via molecular modeling [18,27-33]. Interestingly, coevolving contacts seem to commonly be conserved across distantly related protein pairs: Rodriguez-Rivaz et al. [31] show that contacts predicted in bacterial PPI have a significant chance to be in contact even in eukaryotes.

Despite these successes, the number of analyzed cases remains limited because the coevolution analysis requires a *joint* MSA, with each line containing a pair of interacting proteins. Most protein families contain paralogs (out of 2985 Pfam31 families [9] with more than 1000 sequences from well-defined genomes, 2244 have a mean of more than three paralogs per species, and 1093 a mean above five). Even if we know that the families interact, we generally do not know which paralog from the first family interacts physically with which paralog from the second family. In bacteria, this question can sometimes be answered using the genomic organization in operons: genes coding for interacting proteins are frequently co-localized in an operon. However, taking *Escherichia coli* as a reference, we find 2682 Pfam domain families with at least one hit, corresponding theoretically to almost 3.6 million Pfam family pairs. Only slightly more than 4000 pairs are actually co-localized in operons, but many more are expected to interact even if not co-localized.

The largest-scale assessment of the ability of DCA-type methods to detect inter-protein contacts was therefore performed in the case of homo-oligomers [34], where interaction partners have identical sequences and the matching problem is trivial. The case is, however, complicated by the fact that the distinction between intra- and inter-protein contacts becomes non-trivial when looking to sequence information alone: [34] thus uses monomer structures to identify all those residue pairs with large direct couplings, which are distant in the monomer, and to use them as prediction for potential inter-protein contacts. Note that some residue pairs may be simultaneously tertiary and quaternary contacts: they are excluded by the analysis. Based on almost 2000 distinct protein families, it was found that contacts were faithfully detected for large and well-conserved interfaces, when sufficient sequences are available (typically several hundred sequences of no more than 80% pairwise sequence ID). More involved analyses have to be performed when, e.g., one protein family contains several sub-families with distinct interaction modes [34].

**Coevolutionary analysis identifies specifically interacting protein pairs**

As made clear above, it is a major challenge to generate joint MSA (1) without or with extremely limited operon information – or (2) with substantial but still partial operon co-localization. Case (2) is more tractable: when co-localized proteins are sufficiently numerous, their joint MSA can be used to construct a global coevolutionary model, and subsequently find interacting protein pairs with distant genomic localization using that model. This has been successfully demonstrated for the case of so-called orphan proteins in TCS [35-37].

The most challenging case is one without any co-localization information: can we use the MSA of two families, exclusively, and match sequences computationally such that matched interaction partners are strongly enriched? The basic idea in this context has already been proposed in 2008 [35] within a Bayesian approach: the matching of true interaction partners is expected to provide a strong inter-protein coevolutionary signal. When making errors, this signal is expected to become weaker. If sequences are matched such that the coevolutionary signal is maximized, actual interaction partners should be preferentially paired. A massive problem results from the astronomically large number of possible matchings. Imagine, e.g., two MSA collecting each 1000 sequences from 250 species, with exactly 4 paralogs per species. The number of matchings in each species would be 4! = 24, giving a total number of $24^{250}$ (~$10^{345}$) matchings between the two MSA. For each of them, DCA should be run to estimate the inter-protein coevolutionary signal. Two back-to-back papers [26,38] have now suggested heuristic strategies to progressively match proteins, reaching between 80 and 90% of correctly matched HK and RR in the TCS test case, cf. Fig. 2C.

Importantly, the inter-protein residue-residue contact prediction based on the resulting joint MSA is almost as accurate as for the exact matching, cf. Fig. 2C. While being computationally quite expensive, these approaches seem the most promising way to analyze general pairs of protein families, without requiring complementary information not contained in the MSA (and thus not available for most protein families).

Note that [26,38] concentrate on a few bacterial case studies, where the operon organization can be used as the ground truth. While the results clearly show the potential of the proposed algorithms, a larger-scale analysis beyond sample cases is still missing. In particular eukaryotic proteins might be problematic: Due to the smaller number of sequenced eukaryotic genomes as compared to bacterial ones, large protein families necessarily contain high numbers of paralogs. As a direct consequence, the matching problem becomes much harder.

**Coevolutionary analysis predicts interactions between protein families**

Addressing scale three, it is now clear that coevolutionary models accurately predict inter-protein contacts, when inferred from large MSA joining homologous protein families. Residue pairs with large direct coevolutionary couplings have a high probability to be in contact, cf. e.g. [34,39]. Addressing scale two maximizing the coevolutionary score between two families allows to match specifically interacting proteins, even if these are not known a priori.

This leads back to the scale-one question: Can the existence of large coevolutionary couplings between two protein families be used to predict an interaction between these families? Can protein-protein interaction networks be predicted just from Pfam family alignments? Fig. 3 shows results from [40] for ribosomal proteins: for each subunit, the 10 predictions of highest inter-MSA score are colored – 80% of these pairs are actually in direct contact in the ribosome, as compared to 9% in a random prediction. However, small interfaces are missed, in agreement with the homo-dimer study by [34]. So, while large inter-protein scores can be seen as a good indicator for a family-family interaction, small scores may well appear in interacting pairs, in particular if these have small or not well conserved interfaces.

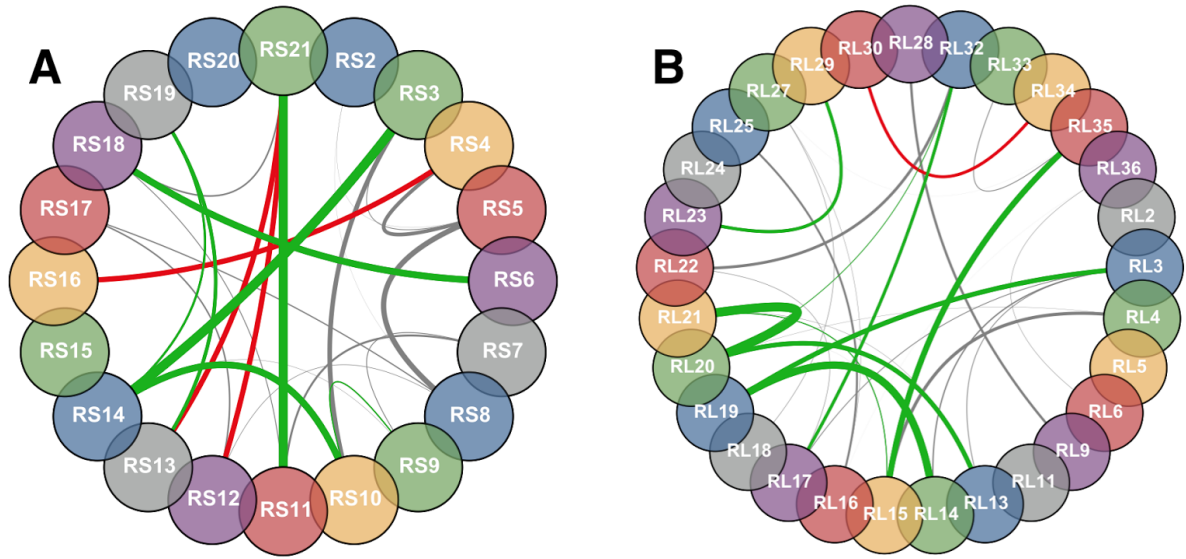

**Figure 3. Coevolutionary prediction of PPI networks for the small (A) and large (B) ribosomal subunit in the bacteria** – Each panel shows the 10 pairs of highest inter-protein coevolutionary score. True positive predictions are drawn in green, false positives in red. Grey lines stand for existing but not predicted interactions. The width of the grey and green lines is proportional to the interface size. We observe that most large interfaces are found, but small ones are typically missed. Figures from [40].

[27] has used this idea to predict family-family interactions in *E. coli*, for all family pairs showing a sufficient number of homologs, which are co-localized in operons, thereby providing a rich source of inspiration for experimental studies. The dimerization of Hsp70, predicted in [29] using DCA, has been functionally confirmed [41].

**Outlook**

We have argued that, within the joint computational framework of global coevolutionary models inferred by DCA, PsiCov or Gremlin, it should be possible to provide purely sequence-based evidence if any two given protein families interact. Forthcoming evidence can be on three different scales, i.e. which families interact, which specific proteins from the families interact, and which residues are in contact – information, which can be used to guide the computational assembly of protein complexes. In this sense, coevolutionary modeling may give important input to what is called "structural systems biology" [42], or what could be called a structurally resolved, conserved interactome.

This objective remains immensely challenging. Pfam currently lists 16,712 protein families [9], corresponding to almost 140 million pairs of families. Even when restricting our analysis to the 2,985 families with currently more than 1,000 sequences (not counting sequences assigned to collections of species at higher taxonomic levels), we would end up with more than 4.4 million family pairs. This number is growing rapidly with each new Pfam release.

Treating this number of family pairs is infeasible with current computational approaches, the heaviest step being the paralog matching procedure. A prior selection of pairs of protein families is needed. In [27], this selection was done, for practical reasons, by the operon co-localization. As an alternative, we have explored the possibility to use phylogenetic profiles [43] [44]. We have observed [45], that the introduction of a global variant of phylogenetic profiling, which uses

"phylogenetic couplings" between protein families instead of correlations in analogy to the direct residue-residue couplings in DCA, strongly improves the detection of previously known inter-family relations (including co-localization and physical PPI). Still, there are hundreds of phylogenetically strongly coupled family pairs without known relation – we may consider them as predictions. Many of them, when applying the aforementioned DCA-based procedure to *E. coli* proteins, show significant coevolution also at the residue level. When applying it to human proteins, the matched MSA become still too small to successfully apply DCA; more sequences and better algorithms are needed.

A second limitation is more intrinsic to current algorithms. DCA-type methods are unsupervised – they analyze an MSA statistically to extract a list of strongly coupled residue pairs, which are expected to be in contact. In the case of tertiary-structure contact prediction, i.e. contacts inside single proteins, an important improvement was obtained by supervised methods [46-48]: the results of DCA are put together with other features, which may include outcomes of other contact-prediction methods, predicted secondary structure or solvent accessibility. Machine learning approaches are used to train predictors on a large number of experimental protein structures. These supervised predictors largely outperform the unsupervised ones; they belong to the most accurate contact predictors to date in the CASP competition (http://predictioncenter.org/). They also partially overcome the strong limitations of global coevolutionary model inference using small MSA. A problem for generalizing this approach to PPI results from the relatively scarce structural knowledge for protein complexes [49] – supervised approaches need large structural training sets to avoid overfitting.

To summarize, recent coevolutionary modeling approaches offer immense opportunities in analyzing protein-protein interactions on multiple scales, ranging from residue-residue contacts up to conserved interactions between protein families. They are based on the statistical exploration of sequence data and will thus fully benefit from the ongoing accumulation of newly sequenced genomes. Integrative approaches adding complementary information to sequences, will further improve the accuracy of coevolutionary models and make them fully valid alternatives to direct experimental approaches. These developments can be expected to make the application of coevolutionary modeling also possible to PPI in higher organisms in the foreseeable future.

**Conflict of interest**

The authors declare no competing financial interests.

**Acknowledgements**

MW acknowledges funding by the Agence Nationale de la Recherche via the project COEVSTAT (ANR-13-BS04-0012-01), and by the European Union's Horizon 2020 research and innovation programme MSCA-RISE-2016 under grant agreement No. 734439 INFERNET. HS was funded by grant GM106085 from the National Institute of General Medical Sciences, National Institutes of Health, USA.